\documentstyle{elsart}

\begin{document}
\begin{frontmatter}
\title{Electrode thickness measurement of a Si(Li) detector for the SIXA
 array}

\author{T.~Tikkanen}
\address{Observatory and Astrophysics Laboratory,
 P.O. Box 14 (T\"ahti\-tornin\-m\"aki),
 FIN-00014 University of Helsinki,
 Finland}
\author{K.~H\"am\"al\"ainen, S.~Huotari}
\address{Department of Physics,
 P.O. Box 9,
 FIN-00014 University of Helsinki,
 Finland}

\begin{abstract}
Cathode electrodes of the Si(Li) detector elements of the SIXA X-ray
 spectrometer array are formed by gold-palladium alloy contact layers.
 The equivalent thickness of gold in one element was measured by
 observing the characteristic L-shell X-rays of gold excited by
 monochromatised synchrotron radiation with photon energies above the
 L$_3$ absorption edge of gold. The results obtained at 4 different
 photon energies below the L$_2$ edge yield an average value of
 $22.4\pm 3.5$~nm which is consistent with the earlier result
 extracted from detection efficiency measurements.

\end{abstract}

\end{frontmatter}

\section{Introduction}

The SIXA (Silicon X-Ray Array) detector \cite{Vilhu} is a focal plane
 instrument of the SODART X-ray telescope on board the Spectrum-X-Gamma
 satellite scheduled for launch in 1998. SIXA detects X-rays in the
 energy range between 500~eV and 20~keV with a closely packed array of
 19 discrete Si(Li) detector elements. The elements are cylindrical and
 their active detection volume is 9.2~mm in diameter and 3.0~mm in
 thickness.

X-rays impinge on the detector through the cathode side of the Si(Li)
 crystals. The cathode consists of a contact layer manufactured by
 sputter deposition of gold-palladium alloy on the crystalline silicon.
 The Au/Pd mass ratio of the alloy is 60:40 and the nominal thickness is
 30~nm but deviations from this value are expected to be large for the
 individual detectors because the Si(Li) crystals are batch processed to
 deposit the alloy. Knowledge of the individual Au/Pd
 thickness for each element in the array is necessary, since the
 electrode thickness is the dominant factor determining the detection
 efficiency in the lower part of the energy range.

The electrode thickness can be deduced from absolute measurements of the
 detection efficiency at low energies. Such a measurement was carried
 out for one detector element using synchrotron radiation in the soft
 X-ray energy region from the electron storage ring BESSY, yielding a
 result of 40.7~nm \cite{Tikkanen}. However, in this case the windowless
 detector was operated in the beamline vacuum, which is not practically
 achievable for all SIXA elements which are assembled in a common
 capsule. Instead, the calibration of the flight detector array will be
 done with harder X-ray sources such as X-ray tubes and radioisotopes.

We used the method of Hansen {\em et al.\ } \cite{Hansen} to measure the
 gold thickness by detecting the gold L-shell fluorescence induced by
 primary X-rays of higher energy. In principle, the palladium thickness
 could be measured similarly using primary photon energies above the
 K edge of palladium, but this measurement would not be possible with
 the flight array because its data acquisition system has an upper limit
 of 20~keV\@. It has been noted that the fluorescence lines are spread
 into broad bands due to the escape of photoelectrons from the contact
 layer into the active detection volume \cite{Maenhaut} and an
 experimental correction for the resulting error was applied by Campbell
 and Wang \cite{Campbell91}. We have modelled the full shape of the
 fluorescence lines, including the low-energy tails as well as the
 high-energy tail contributions from the escape of Auger electrons and
 photoelectrons. Primary photon energies above the L$_1$ absorption edge
 of gold were used in other reported measurements
 \cite{Shima,Maenhaut,Campbell86}, but we used also energies between the
 L$_3$ and L$_2$ edges in order to eliminate the uncertainties from
 quantities related to the individual L subshells.

\section{Experiment}

Using the transfer standard SIXA detector element which has been
 characterised at BESSY in the energy range below 5~keV \cite{Tikkanen},
 we recorded energy spectra using monochromatised synchrotron radiation
 at the beamline X18B of the electron storage ring NSLS at
 the Brookhaven National Laboratory. The detector was operated at about
 106~K in a cryostat sealed with an aluminium-coated polyimide window.
 In addition, an 8.5~$\mu$m thick beryllium window was used to suppress
 light. The windows were transparent in a circular area having a
 diameter of 5.9~mm. Between the detector and the exit window of the
 beamline were an ionisation chamber and about 40~cm of air. We
 attenuated the beam by 8 orders of magnitude in order to reach
 reasonable count rates of about 1000 s$^{-1}$. This was achieved by
 using two pinhole apertures.

Spectra were acquired first with the beam directed at the centre of the
 detector surface. A few examples are shown in Fig.~\ref{fig:data}
 together with model functions of the primary peak including the
 low-energy tail extrapolated from the low-energy characterisation
 results. The gold L$\alpha$ line was best resolvable at the photon
 energy of 12.5~keV and we used this energy when we recorded spectra
 with the beam directed at 4 different positions on the detector
 surface.

\section{Fluorescence model}

When the Si(Li) detector is irradiated with X-rays, the gold atoms in
 the contact layer emit characteristic photons isotropically into all
 directions. The detected fluorescence lines have more tailing than
 lines produced by normally incident primary radiation. High-energy
 tails are formed by events where the characteristic photon is
 accompanied with an energetic electron escaping from the electrode into
 the active detection volume. Low-energy tails due to energy loss across
 the alloy/silicon interface are higher for isotropically incident
 radiation because absorption is more concentrated near the surface.

In the primary photon energy range of 11.919--13.734~keV between the
 L$_3$ and L$_2$ edges, L$\alpha_1$ is the strongest fluorescence line
 and Auger electrons have usually greater energies than photoelectrons.
 The most probable Auger electron to follow L$\alpha_1$ photon emission
 is M$_5$N$_5$N$_7$ (energy 1.79~keV). Its range in the Au/Pd alloy is
 $R_{\mathrm{Aug}}=21$~nm \cite{Fitting}. Assuming isotropic initial
 direction distribution of the Auger electrons, the fraction of Auger
 electrons generated within the distance $d \le R_{\mathrm{Aug}}$ from
 the alloy/silicon interface which will cross the interface is given by
 $p_{\mathrm{iso}}(d/R_{\mathrm{Aug}})$, where
\begin{equation}
p_{\mathrm{iso}}(x) = \quart \arcsin x + \quart x \sqrt{1-x^2}
		    - \frac 1 6 {\left( 2x - x^2 \right)}^\threehalf .
\end{equation}
For most if not all detector elements the electrode thickness $T$ is
 greater than $R_{\mathrm{Aug}}$ and as the primary photons are
 absorbed very homogeneously throughout the whole thickness, the Auger
 electron escape probability is $P_{\mathrm{Aug}} =
 (1 - \omega_{\mathrm M_5}) p_{\mathrm{iso}} R_{\mathrm{Aug}}/T$, where
 $\omega_{\mathrm M_5}$ is the M$_5$-shell fluorescence yield and
 $p_{\mathrm{iso}} \equiv p_{\mathrm{iso}}(1)=\frac{\pi}{8}-\frac 1 6$.
 The photoelectron escape probability is obtained similarly:
 $P_{\mathrm{pho}} = p_{\mathrm{pho}} R_{\mathrm{pho}}/T$, where
 $p_{\mathrm{pho}}=0.153$ \cite{Scholze} is the average escape
 probability of the photoelectrons generated within their range
 $R_{\mathrm{pho}}$ from the interface. Adding these and subtracting
 the probability of simultaneous escape of both electrons to prevent
 these events from being counted twice, the tail-to-total ratio for the
 high-energy tail is obtained as
\begin{equation}
\label{eq:high}
P_{\mathrm h}
  = P_{\mathrm{Aug}} + P_{\mathrm{pho}} (1 - P_{\mathrm{Aug}}) .
\end{equation}

The components of the low-energy tail for isotropically incident
 radiation are obtained by integration over all directions. For
 the silicon escape peak the result is, neglecting
 escape through the sides and rear of the cylinder,
\begin{equation}
P_{\mathrm{esc}} = \frac 1 8 \omega_{\mathrm K} f_{\mathrm K}
 \left[ 1 - x \ln \left( 1 + \frac 1 x \right)
	  + \frac 1 x \ln \left( 1 + x \right) \right] ,
\end{equation}
where $\omega_{\mathrm K}$ is the K-shell fluorescence yield of silicon,
 $f_{\mathrm K}$ is the fractional contribution of the K shell to $\mu$
 (the linear attenuation coefficient of silicon at the primary photon
 energy) and $x=\mu_{\mathrm K}/\mu$, where $\mu_{\mathrm K}$ is the
 linear attenuation coefficient of silicon at the K line energy. The
 other components are derived from the physical model of Scholze and Ulm
 \cite{Scholze}. For the short tail, the integration yields
\begin{equation}
P_{\mathrm t} = \half \int_0^1 \left( 1 - \e^{-\mu R/x} \right) \d x ,
\end{equation}
where $R=210$~nm is a parameter of the model. When calculating the
 escape of the primary electrons from silicon into the electrode, it is
 assumed that the photoelectrons are ejected isotropically into all
 directions because the absorbed photons arrive isotropically from the
 half-space. The resulting contribution to the flat shelf is
\begin{equation}
\label{eq:si}
P_{\mathrm{Si}} = \half p_{\mathrm{iso}}
	\int_0^1 \left( 2 - \e^{-\mu R_{\mathrm{pho}}/x}
			  - \e^{-\mu R_{\mathrm{Aug}}/x} \right) \d x .
\end{equation}
The low-energy tail contributions are compared to those for normally
 incident radiation from an external source
 of the same energy in Table~\ref{tab:tail}.
 The flat shelf contribution was calculated with $T=40.7$~nm. The total
 fraction of counts in the low-energy tail is 1.8\% while for normally
 incident radiation it is 0.7\%.

When the primary beam is directed at the central region, all L-series
 characteristic photons emitted by gold atoms into the half-space formed
 by the Si(Li) bulk are detected and the full-peak absolute quantum
 detection efficiency for the characteristic line is
\begin{equation}
 \epsilon_{\mathrm{Au}} = \half \left( 1 - A \right)
 \left( 1 - P_{\mathrm h} \right)
 \left( 1 - P_{\mathrm{esc}} - P_{\mathrm t} - P_{\mathrm{Si}} \right) ,
\end{equation}
where $A$ is the self-absorption probability in the electrode given by
\begin{equation}
A = 1 - \int_0^1 \frac{x}{\mu_{\mathrm{Au/Pd}} T}
		 \left( 1 - \e^{-\mu_{\mathrm{Au/Pd}} T/x} \right)
		 \d x ,
\end{equation}
which yields 2.2\% for the L$\alpha$ line.
 With the approximation that the self-absorption events are uniformly
 distributed across the electrode thickness, the shelf contribution from
 the escape of primary electrons into the active detection volume is
 obtained as
\begin{equation}
\label{eq:aupd}
P_{\mathrm{Au/Pd}}
 = \frac{A}{\mu_{\mathrm{Au}} + \mu_{\mathrm{Pd}}}
   \sum_{i=\mathrm{Au}}^{\mathrm{Pd}} \mu_i
      \left[ p_{\mathrm{iso}}\left( \frac{T}{R_{\mathrm{pho},i}} \right)
	   + p_{\mathrm{iso}} \frac{R_{\mathrm{Aug},i}}{T} \right] ,
\end{equation}
since the applicable photoelectron ranges are now greater than $T$.

Calculated shapes of the fluorescence lines from L$_3$ shell ionisation
 are shown in Fig.~\ref{fig:shape}, including L$\alpha_1$,
 L$\beta_{2,15}$, L$\alpha_2$, L$\iota$, L$\beta_5$ and L$\beta_6$.
 The electron ranges in Eqs.~(\ref{eq:high}), (\ref{eq:si}) and
 (\ref{eq:aupd}) were calculated using the most probable Auger and
 photoelectron energies. The high-energy tail, being produced like the
 flat shelf by escape of energetic electrons, consists of two step
 functions convolved with a Gaussian, one due to the Auger electron and
 the other one to the photoelectron. For primary photon energies above
 the L$_2$ edge, a third one due to a Coster--Kronig electron is added.

When the primary photon energy $E$ is between the ionisation energies of
 the L$_3$ and L$_2$ shells of gold, the intensity ratio of the Au
 L$\alpha$ line and the primary line is given by
\begin{equation}
\frac{I_{\mathrm{Au}}}{I(E)}
 = \omega_{\mathrm L_3} f_{\mathrm L\alpha}
   \frac{\epsilon_{\mathrm{Au}}}{\epsilon(E)}
   \left( 1 - \e^{-\mu_{\mathrm L_3}(E) t} \right) ,
\end{equation}
where $\omega_{\mathrm L_3}$ is the fluorescence yield of the gold L$_3$
 shell, $f_{\mathrm L\alpha}$ is the intensity ratio of L$\alpha$ to all
 L$_3$-series lines, $\epsilon$ is the quantum detection efficiency of
 the Si(Li) detector, $\mu_{\mathrm L_3}$ is the L$_3$-shell linear
 attenuation coefficient of gold and $t$ is the equivalent gold
 thickness. Above the L$_2$ and L$_1$ absorption edges the L$\alpha$
 line includes contributions from absorptions to the other L subshells
 followed by Coster--Kronig transitions. With the approximation
 $1 - \e^{-\mu t} \approx \mu t$ the thickness is given by
\begin{equation}\label{eq:thick}
t = \left( \frac{I(E)}{I_{\mathrm{Au}}} \right)
    \left( \frac{1}{\omega_{\mathrm L_3} f_{\mathrm L\alpha}} \right)
    \frac{\epsilon(E)}
	 {\epsilon_{\mathrm{Au}} \mu_{\mathrm L_3}(E)
	  + f_{23} \epsilon_{\mathrm{Au},2} \mu_{\mathrm L_2}(E)
	  + \left( f_{13} + f_{12} f_{23} \right)
	    \epsilon_{\mathrm{Au},1} \mu_{\mathrm L_1}(E)} ,
\end{equation}
where the $f$'s are the applicable Coster--Kronig yields and
 $\epsilon_{\mathrm{Au},2}$ and $\epsilon_{\mathrm{Au},1}$ have
 different high-energy tails than $\epsilon_{\mathrm{Au}}$ because the
 photoelectron energies are different and the escape of Coster--Kronig
 electrons add new tail contributions. Our calculation includes the
 high-energy tail contribution
 $P_{\mathrm{CK},2} = \half f_{23} p_{\mathrm{iso}} R_{23}/T$ for
 $\epsilon_{\mathrm{Au},2}$, where the factor $\half$ represents roughly
 the probability that the Coster--Kronig electron is energetic (i.e.\ 
not of the LLM series) and $R_{23}$ is the range of the L$_2$L$_3$N$_1$
 Coster--Kronig electron. For $\epsilon_{\mathrm{Au},1}$ we set
 $P_{\mathrm{CK},1} = f_{13} p_{\mathrm{iso}} R_{13}/T$. The
 contributions of the individual L subshells to $\mu$ were calculated
 from the jump ratios using the data of Henke {\em et al.\ }
 \cite{Henke}.

\section{Results}

Net counts in the gold fluorescence lines were calculated by fitting the
 modelled shape of the lines plus continuum to the data in the region
 of the L$\alpha$ peak. Ideally, the continuum is formed by the
 low-energy tail of the primary peak. However, the measured continua are
 higher than the modelled tails in certain energy regions (see
 Fig.~\ref{fig:data}). A broad feature above 8~keV appears in all
 spectra up to the primary photon energy of 13.0~keV, including the data
 taken at 11.8~keV which is below the L edges of gold. This feature can
 be a measurement artifact or reflect the real shape of the tail; our
 extrapolated tail model is based on a physical model which applies to
 photon energies below 4~keV \cite{Scholze}. At photon energies above
 13.0~keV, the L$\alpha$ line of lead at 10.5~keV appears. This line
 originates in lead impurities in the slits and apertures. The continuum
 might as well be modelled using the original form of the modified
 HYPERMET function \cite{Phillips,Campbell85}. We tried both forms and
 added a first-order polynomial to fit the unknown feature. The best-fit
 lead L$\alpha$ lines were added to the continuum at photon energies
 above the L$_3$ absorption edge of lead. An example is shown in
 Fig.~\ref{fig:fit} (original form).

Fig.~\ref{fig:thick} presents the thicknesses obtained from
 Eq.~(\ref{eq:thick}), both using the original HYPERMET form and the
 form of Ref.~\cite{Scholze}. Statistical uncertainties (standard
 deviation in the true photon number) were derived from the net counts
 in the L$\alpha$ peak. Weighted average of the results from 12.0 to
 13.5~keV at the detector centre is 22.4~nm with a standard deviation of
 3.5~nm with the original form and $22.2\pm 3.5$~nm with the other form.
 The 16\% standard deviations are higher than the real uncertainty
 of the intensity. This is explained by the fluctuations of the
 continuum counts and by errors in the electron ranges, absorption
 coefficients and other data. Inclusion of the results up to 16.0~keV
 yields $22.8\pm 2.9$~nm with the original form (results at 17.0~keV and
 above are excluded because the continuum is dominated by Compton
 scattering). The results at 12.5~keV at different positions yield an
 average of 23.9~nm with a standard deviation of 2.3~nm which means that
 thickness variations across the detector surface greater than the
 overall precision were not found.

\section{Conclusion}

The gold-palladium alloy contact layer thickness of a Si(Li) detector
 element for the SIXA array was measured using monochromatised
 synchrotron radiation to excite L-shell fluorescence from gold. The
 fluctuation of the result as function of primary photon energy yields
 an accuracy of about 3~nm in the equivalent thickness of gold. The
 result is in agreement with an earlier result $20.1\pm 0.3$~nm
 derived from a detection efficiency measurement \cite{Tikkanen}.
 The contact layer thicknesses of the elements in the SIXA flight array
 are to be measured using monochromatised radiation of an X-ray tube
 with longer acquisition times which will reduce the uncertainty from
 photon statistics. The accuracy will be limited by the knowledge of the
 shape of the low-energy tail of the primary peak and the uncertainty in
 the electron ranges and the absorption coefficients and other data
 regarding the L subshells of gold.

\begin{ack}
We thank L. Furenlid of the Brookhaven National Laboratory for
collaboration at the beamline and
 Metorex International Oy (Espoo, Finland) for providing the detector.
S. Kraft of the Physikalisch-Technische Bundesanstalt, Berlin, is
acknowledged for extrapolating the HYPERMET function parameters to
the higher part of our energy range.
This work was supported by the Academy of Finland (contract SA-8582).
\end{ack}

\newpage

\begin{figure}[h]
\caption{Measured pulse height spectra of monochromatised X-rays at
 photon energies 11.8, 12.5, 13.5 and 15.0~keV\@. The curves show the
 calculated shape of the primary line including its low-energy tail.}
\label{fig:data}
\end{figure}

\begin{figure}[h]
\caption{Calculated pulse height spectra of X-ray fluorescence from the
 electrode with primary photon energies of 12.0~keV (dashed curve),
 13.5~keV (solid curve) and 15.0~keV (dotted curve). The total count
 rate is normalised to the total rate of primary photon absorption
 events by gold atoms.}
\label{fig:shape}
\end{figure}

\begin{figure}[h]
\caption{Illustration of the data analysis at the primary photon energy
 of 13.0~keV: measured data (points) and the best-fit model function
 (curve) are shown in the upper panel and their difference in the energy
 region used in the fitting in the lower panel. The thin curves are the
 gold L$\alpha$ fluorescence line and the added linear background
 component.}
\label{fig:fit}
\end{figure}

\begin{figure}[h]
\caption{Equivalent gold thickness calculated from data obtained using
 different primary photon energies. The error bars show the statistical
 $1\sigma$ uncertainties from the limited number of the gold L$\alpha$
 photons. Results using an alternative method to compute net counts
 (extrapolated low-energy model function for the tail of the primary
 peak) are indicated by squares. Off-centre results acquired at 12.5~keV
 are plotted at 12.4~keV for the sake of clarity.}
\label{fig:thick}
\end{figure}

\newpage

\begin{table}[bht]
\caption{Components of the low-energy tail of the Si(Li) detector
 for the L$\alpha_1$ characteristic line of gold: normally incident
 radiation (left column), fluorescence generated in the electrode
 (right column).}
\label{tab:tail}
\begin{tabular}{|l|c|c|}
\multicolumn{3}{c}{ } \\ \hline
 & External (\%) & Electrode (\%) \\ \hline
Si escape & 0.10 &  0.16 \\
Short tail & 0.17 & 0.58 \\
Flat shelf (Si) & 0.10 & 0.41 \\
Flat shelf (total) & 0.46 & 1.01 \\
 \hline
\end{tabular}
\end{table}

\end{document}